\documentclass[11pt]{article}

\usepackage[T1]{fontenc}
\usepackage[sc]{mathpazo}
\usepackage{amsmath}
\usepackage{amssymb}
\usepackage{enumerate}
\usepackage{amsthm}
\usepackage{amsfonts,mathrsfs}
\usepackage[style]{fncychap}
\usepackage{graphicx} 
\usepackage{geometry} 
\usepackage{eepic}
\usepackage{ifthen}
\newboolean{ElectronicVersion}
\setboolean{ElectronicVersion}{true} 

\usepackage{hyperref}
\hypersetup{pdfpagemode=UseNone}

\newcommand{\setft}[1]{\mathrm{#1}}
\newcommand{\lin}[1]{\setft{L}\left(#1\right)}

\newenvironment{mylist}[1]{\begin{list}{}{
    \setlength{\leftmargin}{#1}
    \setlength{\rightmargin}{0mm}
    \setlength{\labelsep}{2mm}
    \setlength{\labelwidth}{8mm}
    \setlength{\itemsep}{0mm}}}
    {\end{list}}

\def\ot{\otimes}

\newcommand{\out}[2]{| #1\rangle\langle #2 |}

\newcommand{\Pa}[1]{\left(#1\right)}

\newcommand{\bra}[1]{\langle#1|}

\newcommand{\ket}[1]{|#1\rangle}

\DeclareMathOperator{\trace}{Tr}

\newcommand{\Ptr}[2]{\trace_{#1}\Pa{#2}}

\newcommand{\Tr}[1]{\Ptr{}{#1}}

\def\cC{\mathcal{C}}
\def\cH{\mathcal{H}}
\def\cK{\mathcal{K}}\def\cM{\mathcal{M}}

\def\cV{\mathcal{V}}
\def\cZ{\mathcal{Z}}

\def\A{\textsf{A}}\def\B{\textsf{B}}\def\D{\textsf{D}}

\def\T{\textsf{T}}

\newtheorem{thrm}{Theorem}[section]
\newtheorem{lem}[thrm]{Lemma}
\newtheorem{prop}[thrm]{Proposition}

\theoremstyle{definition}
\newtheorem{definition}[thrm]{Definition}

\newtheorem{exam}[thrm]{Example}
\numberwithin{equation}{section}

\newcounter{questionnumber}

\begin{document}

\title{\Large The Local Orthogonality between Quantum States and Entanglement Decomposition}

\author{Sunho Kim,\,\, Junde Wu \\{\small \it Department of Mathematics, Zhejiang
University, Hangzhou 310027, PR~China}\\ Lin Zhang\\{\small \it
Institute of Mathematics, Hangzhou Dianzi University, Hangzhou
310018, PR~China}\\ Minhyung Cho\\{\small \it Department of Applied
Mathematics, Kumoh National Institute of Technology, Kyungbuk,
730-701, Korea}}

\date{}
\maketitle \mbox{}\hrule\mbox\\
\begin{abstract}
In the paper, we show that when a quantum state can be decomposed as
a convex combination of locally orthogonal mixed states, its
entanglement can be decomposed into the entanglement of these mixed
states without losing them. The obtained result generalizes a
corresponding one proved by Horodecki [Acta Phys. Slov. 48, 141
(1998).]. But, for the entanglement cost it requires certain
conditions for holding the decomposition, and the distillable
entanglement only has a week result as inequality. Finally, we
presented an example to show that the conditions of our conclusions
are existence.
\end{abstract}
\mbox{}\hrule\mbox\\


{\bf The corresponding author}: Minhyung Cho, E-mail:
chominhyung@126.com

\section{Introduction and preliminaries}

In this paper, we always assume that $\cH_A$,  $\cH_B$, $\cK_A$ and
$\cK_B$ are finite dimensional complex Hilbert spaces. Let
$\lin{\cH_A, \cK_A}$ be the set of all linear operators from $\cH_A$
to $\cK_A$. A quantum state $\rho$ of some quantum system, described by
$\cH_A$, is a positive semi-definite operator of trace one, in
particular, for each unit vector $\ket{\psi} \in \cH_A$, the
operator $\rho = \out{\psi}{\psi}$ is said to be a \emph{pure
state}. We can identify the pure state  $\out{\psi}{\psi}$ with the
unit vector $\ket{\psi}$. The set of all quantum states on $\cH_A$ is
denoted by $\D(\cH_A)$.

For each quantum state $\rho\in\D(\cH_A)$, its \emph{von Neumann
entropy} is defined by
$$
S(\rho) = - Tr{\rho\log_2\rho}.
$$

Let $p = (p_{a})\in \mathbb{R}^{\Sigma}$ be a probability distribution, the \emph{Shannon entropy} $H(p)$ of $p$  is defined by
$$
H(p) = -\sum_{a\in\Sigma}p_{a}\log_2 p_{a}.
$$

For given probability distribution $p = (p_{a})\in
\mathbb{R}^{\Sigma}$, positive integer number $n$ and $\varepsilon
>0$, we say that a string $a_{1}\cdots
a_{n}\in\Sigma^{n}=\Sigma\times\Sigma\cdots\Sigma$ is
$\varepsilon$\emph{-typical} if $$
2^{-n(H(p)+\varepsilon)}<p_{a_{1}\cdots
a_{n}}<2^{-n(H(p)-\varepsilon)},$$ where $p_{a_{1}\cdots a_{n}} =
p_{a_{1}}\cdots p_{a_{n}}$. The set of all $\varepsilon$-typical
strings is denoted by $T_{n, \varepsilon}$, that is,
\begin{eqnarray*}
T_{n, \varepsilon} = \{a_{1}\cdots a_{n}\in\Sigma^{n}:
2^{-n(H(p)+\varepsilon)}<p_{a_{1}\cdots
a_{n}}<2^{-n(H(p)-\varepsilon)}\}.
\end{eqnarray*}

The $\varepsilon$-typical string set $T_{n, \varepsilon}$ has the
following property \cite{Watrous}:
\begin{eqnarray*}
\lim_{n\rightarrow\infty}\sum_{a_{1}\cdots a_{n}\in T_{n,
\varepsilon}}p_{a_{1}\cdots a_{n}} = 1.
\end{eqnarray*}

For each $t = a_{1}\cdots a_{n}\in\Sigma^n$, we denote $n_{a,t}$ is
the times of $a$ appearing in $t$. If $a$ does not appear in $t$, we
denote $n_{a,t} = 0$. It is clear that $\sum_{a\in \Sigma}n_{a,t} =
n$. Moreover, we say that a set $T_{\varepsilon}^{n}$ is
$\varepsilon$\emph{-strong typical set} \cite{Hayden}, if
\begin{eqnarray*}
T_{\varepsilon}^{n} = \{s \in \Sigma^{n} : p_{s} = \prod_{a\in
\Sigma}p_{a}^{n_{a,s}},\ n_{a,s}\in[p_{a}n-\frac{\varepsilon
n\log_{p_{a}}2}{|\Sigma|}, p_{a}n+\frac{\varepsilon
n\log_{p_{a}}2}{|\Sigma|}]\}.
\end{eqnarray*}
We can easily see that $T_{\varepsilon}^{n} \subseteq T_{n,
\varepsilon},$ and similar to $\varepsilon$-typical set we have the
following property \cite{Hayden}, for complete sake, we prove them.
\begin{lem}
Let $p = (p_{a})\in \mathbb{R}^{\Sigma}$ be a probability
distribution and let $\varepsilon > 0,$ then
\begin{eqnarray*}
\lim_{n\rightarrow\infty}\sum_{s\in T_{\varepsilon}^{n}}p_{s} = 1.
\end{eqnarray*}
\end{lem}

\begin{proof}
Let $X_{a}^{n}, X_{b}^{n}, \cdots$ be independent for each $a, b,
\cdots \in \Sigma$ and positive integer $n$. The random variables
are defined as follows: for each $a\in \Sigma$ randomly according to
the probability distribution $p$, let $X_{a}^{n}$ be the times of
$p_{a}$ appears in $p_{a_{1}\cdots a_{n}}$. It holds that
\begin{eqnarray*}
\frac{X_{a}^{n} - np_{a}}{\sqrt{np_{a}(1-p_{a})}} \sim N(0, 1) \ \
(approximate),
\end{eqnarray*}
when $n\rightarrow\infty.$

Therefore, for $\varepsilon>0,$
\begin{eqnarray*}
&&Pr[p_{a}n-\frac{\varepsilon n\log_{p_{a}}2}{|\Sigma|} \leq
X_{a}^{n} \leq p_{a}n+\frac{\varepsilon n\log_{p_{a}}2}{|\Sigma|}]
\\ = &&Pr[-\frac{\varepsilon
n\log_{p_{a}}2}{\sqrt{np_{a}(1-p_{a})}|\Sigma|} \leq \frac{X_{a}^{n}
- np_{a}}{\sqrt{np_{a}(1-p_{a})}} \leq \frac{\varepsilon
n\log_{p_{a}}2}{\sqrt{np_{a}(1-p_{a})}|\Sigma|}],
\end{eqnarray*}
and $\frac{\varepsilon
n\log_{p_{a}}2}{\sqrt{np_{a}(1-p_{a})}|\Sigma|} \rightarrow\infty$
as $n\rightarrow\infty,$ so we have
\begin{eqnarray*}
\lim_{n\rightarrow\infty}Pr[p_{a}n-\frac{\varepsilon
n\log_{p_{a}}2}{|\Sigma|} \leq X_{a}^{n} \leq
p_{a}n+\frac{\varepsilon n\log_{p_{a}}2}{|\Sigma|}] = 1.
\end{eqnarray*}
Note that the random variables $X_{a}^{n}, X_{b}^{n}, \cdots$ are
independent for each $a, b, \cdots \in \Sigma$, we have
\begin{eqnarray*}
\lim_{n\rightarrow\infty}\sum_{s\in T_{\varepsilon}^{n}}p_{s} = 1.
\end{eqnarray*}
\end{proof}

\begin{lem}\label{lem:lem-1}
Let $\rho\in\D(\cH_A)$ be composed of quantum state ensemble
$\{\rho_{a}\}_{a\in\Sigma}$ with probability
distribution $p=(p_{a})$ such that $\rho =
\sum_{a\in\Sigma}p_{a}\rho_{a}$. For each
$t=a_{1}\cdots a_{n}\in\Sigma^n$, if we denote
$p_t=p_{a_{1}}\cdots p_{a_{n}}$,
$\rho_t=\rho_{a_1}\ot\rho_{a_2}\ot\cdots\ot\rho_{a_n}$, and
$\rho_{T_{\varepsilon}^{n}} = \sum_{s\in T_{\varepsilon}^{n}}p_{s}
\rho_{s}$, then $$\lim_{n\rightarrow\infty}\|\rho^{\otimes
n}-\rho_{T_{\varepsilon}^{n}}\|_{1}=0.$$
\end{lem}

\begin{proof} Note that the quantum state $\rho^{\otimes n}$ can be
decomposed into
\begin{eqnarray*}
\rho^{\otimes n} = \sum_{t\in
\Sigma^{n}}\prod_{a\in \Sigma}p_{a}^{n_{a,t}}
\rho_t.
\end{eqnarray*}
If we denote \begin{eqnarray*} \rho_{T_{\varepsilon}^{n}} =
\sum_{s\in T_{\varepsilon}^{n}}\prod_{a\in \Sigma}p_{a}^{n_{a,s}}
\rho_s,
\end{eqnarray*}
then
\begin{eqnarray*}
\lim_{n\rightarrow\infty}\sum_{t\in \Sigma^{n}\setminus
T_{\varepsilon}^{n}}\prod_{a\in \Sigma}p_{a}^{n_{a,t}} = 0.
\end{eqnarray*}
Therefore
\begin{eqnarray*}
\lim_{n\rightarrow\infty}\|\rho^{\otimes
n}-\rho_{T_{\varepsilon}^{n}}\|_{1} &=&
\lim_{n\rightarrow\infty}\|\sum_{t\in \Sigma^{n}\setminus
T_{\varepsilon}^{n}}\prod_{a\in \Sigma}p_{a}^{n_{a,t}}
\rho_{t}\|\\
&\leq& \lim_{n\rightarrow\infty}\sum_{t\in \Sigma^{n}\setminus
T_{\varepsilon}^{n}}\prod_{a\in \Sigma}p_{a}^{n_{a,t}} = 0.
\end{eqnarray*}
\end{proof}

The \emph{fidelity} between two quantum states $\rho$ and $\sigma$
is defined by
\begin{eqnarray*}
F(\rho, \sigma ) = Tr{\sqrt{\sqrt{\rho}\sigma\sqrt{\rho}}}.
\end{eqnarray*}

Let $\out{\phi}{\phi}\in\D(\cH_A)$ be a pure state and $\rho =
\sum_{a\in\Sigma}p_{a}\out{\psi_{a}}{\psi_{a}}\in\D(\cH_A)$ be any
quantum state. Note that for each $\lambda\geq0$,
$\sqrt{\lambda\out{\phi}{\phi}} = \sqrt{\lambda}\out{\phi}{\phi}$,
so
\begin{eqnarray*}
(F(\out{\phi}{\phi}, \rho))^{2} = \langle\out{\phi}{\phi},
\rho\rangle = \sum_{a\in\Sigma}p_{a}\langle\out{\phi}{\phi},
\out{\psi_{a}}{\psi_{a}}\rangle =
\sum_{a\in\Sigma}p_{a}(F(\out{\phi}{\phi},
\out{\psi_{a}}{\psi_{a}}))^{2},
\end{eqnarray*}
in particular, $F(\out{\phi}{\phi}, \out{\psi}{\psi}) =
|\langle\phi, \psi\rangle|$ for any unit vectors $\ket{\phi},
\ket{\psi} \in\cH_A$.

 Let  $\rho\in\D(\cH_A)$. \emph{A purification} of
$\rho$ in $\cH_A\ot\cH_B$ is any pure state $\out{u}{u}$ in
$\cH_A\ot\cH_B$ for which the partial trace $Tr_{\cH_B}(\out{u}{u})$
of $\out{u}{u}$ over subsystem $\cH_B$ is $\rho$.

Let $\T(\cH_A, \cK_A)$ denote the set of all \emph{linear
super-operators} from $\lin{\cH_A}$ to $\lin{\cK_A}$. We say that
$\Phi\in\T(\cH_A, \cK_A)$ is \emph{completely positive} if for each
positive integer $k\in\mathbf{N}$, $\Phi\ot{M_{k}}:\lin{\cH}\ot
M_{k}\to\lin{\cH}\ot M_{k}$ is positive, where $M_{k}$ is the set of
all $k\times k$ complex matrices. It follows from the famous
theorems of Choi \cite{Choi} and Kraus \cite{Kraus1} that if $\Phi$
is complete positive, then it can be represented in the following
form $$\Phi(X)=\sum_{j=1}^nM_jXM_j^\dagger,\,\, X\in \lin{\cH_A},$$ where $\{M_j\}_{j=1}^n\subseteq \lin{\cH_A, \cK_A}$,
$M^\dagger$ is the \emph{adjoint operator} of $M$. In this case, we
denote $\Phi=\sum_{j}Ad_{M_{j}}$. If $\sum_\mu M^\dagger_\mu M_\mu =
I_{\cH_A}$, then $\Phi=\sum_{j}Ad_{M_{j}}$ is said to be \emph{an
admissible quantum operation}.

Let $\Phi_A\in T(\cH_A, \cK_A)$,  $\Phi_B\in T(\cH_B, \cK_B)$ be two admissible quantum operations. Then  $\Phi_A\ot\Phi_B\in T(\cH_A\ot\cH_B, \cK_A\ot\cH_B)$ is said to be \emph{an admissible product quantum operation}.

Let $\cZ_A$ be a finite dimensional complex Hilbert space, $\Sigma$
be a finite set,
$$\{P_a: a\in \Sigma\}$$ be a measurement on $\cZ_A$, and let
$\cZ_B=\mathbf{C}^{\Sigma}$, $\{e_a: a\in\Sigma\}$ be the standard
orthogonal basis of $\cZ_B$, $E_{a,a}=\out{e_a}{e_a}$. The
super-operator of the form $$\Phi\in T(\cH_A\ot\cZ_A\ot\cH_B,
\cH_A\ot\cH_B\ot\cZ_B)$$ defined by
$$\Phi(X)=\sum_{a\in\Sigma}Tr_{\cZ_A}[(I_{\cH_A}\ot P_a\ot I_{\cH_B})X]\ot E_{a,a}$$
is an \emph{Alice-Bob measurement transmission super-operator}.

The interpretation of such a super-operator is that Alice performs
the measurement described by $\{P_a: a\in\Sigma\}$ on the part of her
quantum system $\cZ_A$ and transmits the result to Bob. We then
imagine that Bob initializes a quantum register to the state
$E_{a,a}$ for whichever outcome $a\in\Sigma$ Alice obtained, so that
we may incorporate this measurement outcome into the description of
Bob's quantum information.

A \emph{Bob-to-Alice measurement transmission super-operator} is a
super-operator of the form
$$\Phi\in T(\cH_A\ot\cH_B\ot\cZ_B, \cH_A\ot\cZ_A\ot\cH_B)$$ that is defined in the same way as an Alice-Bob measurement transmission super-operator, except that of course Bob performs the measurement rather than
Alice.

We may speak of a \emph{measurement transmission super-operator} to
mean either an Alice-to-Bob or Bob-to-Alice measurement transmission
super-operator.

An \emph{LOCC} super-operator is any super-operator of the form
$$\Phi\in T(\cH_A\ot\cH_B, \cK_A\ot\cK_B)$$ that can
be obtained from the composition of any finite number of admissible
product super-operators and measurement transmission
super-operators.

We will write $$LOCC(\cH_A, \cK_A: \cH_B, \cK_B)$$ to denote the
collection of all LOCC super-operators as just defined. For much
more notations and definitions, see \cite{Watrous}.

Given a quantum state $\rho\in \D(\cH_{A}\ot \cH_{B})$, denote
$\rho_{A}=Tr_{\cH_B}\rho$, and $\rho_{B}=Tr_{\cH_A}\rho$,
respectively. If $\rho$ is the pure state $\out{\psi}{\psi}$ in
$\D(\cH_{A}\ot \cH_{B})$, then $S(\rho_{A})=S(\rho_{B})$
\cite{Bennett}. The entanglement $E(\ket{\psi})$ of pure state
$\out{\psi}{\psi}$ is defined by
$$E(\ket{\psi}) = S(\rho_{A}) = S(\rho_{B}).$$

Given a quantum state $\rho\in D(\cH_{A}\otimes \cH_{B})$, consider
all possible pure state ensemble $\{\out{\psi_{a}}{\psi_{a}}\}_{a\in\Sigma}$ with
probability distribution $p=(p_{a})$ such that $\rho =
\sum_{a\in\Sigma}p_{a}\out{\psi_{a}}{\psi_{a}}$. The \emph{entanglement of
formation $E_{f}(\rho)$} of $\rho$ is defined by (\cite{{Bennett}},
\cite{Wootters}):
\begin{eqnarray*}
E_{f}(\rho) = \min_{\Sigma} \sum_{a\in\Sigma}p_{a}E(\ket{\psi_{a}}).
\end{eqnarray*}

Let $\tau\in D(\cC^{\{0,1\}}\ot\cC^{\{0,1\}})$ denote the operator
\begin{eqnarray*}
\tau = \frac{1}{2}(\ket{00}+\ket{11})(\bra{00}+\bra{11}).
\end{eqnarray*}

The \emph{entanglement cost $E_{c}(\rho)$} of $\rho\in
D(\cH_{A}\otimes \cH_{B})$ is
the infimum over all real numbers $\alpha\geq 0$ (\cite{Hayden}, \cite{Bennett}), for which there
exists a sequence of LOCC super-operators
\begin{eqnarray*}
\Phi_{n}\in LOCC((\cC^{\{0,1\}})^{\ot\lfloor \alpha
n\rfloor},\cH_{A}^{\ot n} : (\cC^{\{0,1\}})^{\ot\lfloor \alpha
n\rfloor},\cH_{B}^{\ot n}),
\end{eqnarray*}
such that $\lim_{n\rightarrow\infty}F(\Phi_{n}(\tau^{\ot\lfloor \alpha n\rfloor}), \rho^{\otimes n}) = 1.$

The \emph{distillable entanglement $E_{d}(\rho)$} of a quantum state
$\rho\in D(\cH_{A}\otimes \cH_{B})$ is the supremum over all real
numbers $\alpha\geq 0$ (\cite{Bennett}, \cite{Rains}), for which there exists a sequence of LOCC
super-operators
\begin{eqnarray*}
\Phi_{n}\in LOCC(\cH_{A}^{\ot n}, (\cC^{\{0,1\}})^{\ot\lfloor \alpha
n\rfloor} : \cH_{B}^{\ot n}, (\cC^{\{0,1\}})^{\ot\lfloor \alpha
n\rfloor}),
\end{eqnarray*}
such that $\lim_{n\rightarrow\infty}F(\Phi_{n}(\rho^{\otimes n}), \tau^{\ot\lfloor \alpha n\rfloor}) = 1.$

For each $\rho, \sigma\in D(\cH_{A}\otimes \cH_{B})$, we have the
following important facts (\cite{Hayden}, \cite{Vidal-1}, \cite{Bennett-2}):

(1). $E_{f}(\rho \otimes \sigma) \leq E_{f}(\rho) + E_{f}(\sigma)$.

(2). For any family of \ CP local maps \ $\mathcal{M}^{(i)}_{loc}(\rho) = \sum_{j,k}\A_{ij}\otimes\B_{ik} \ \rho \ \A_{ij}^{\dagger}\otimes\B_{ik}^{\dagger}$ \ such that \ $\sum_{i,j,k}\A_{ij}^{\dagger}\A_{ij}\otimes\B_{ik}^{\dagger}\B_{ik} = 1$.
\ Then $E_{f}(\rho)$ satisfies the monotonicity condition
\begin{eqnarray*}
\sum_{i} p_{i}E_{f}(p_{i}^{-1}\mathcal{M}^{(i)}_{loc}(\rho)) \leq E_{f}(\rho), \ \ with \ \ p_{i} = Tr[\mathcal{M}^{(i)}_{loc}(\rho)].
\end{eqnarray*}

(3). $E_{d}(\rho) \leq E_{c}(\rho) \leq E_{f}(\rho)$.

(4). If $\rho$ is a pure state $\out{\psi}{\psi}$, then
$$E_{f}(\out{\psi}{\psi}) = E_{c}(\out{\psi}{\psi}) =
E_{d}(\out{\psi}{\psi}).$$

(5). $E_{c}(\rho) = \lim_{n\rightarrow\infty}
\frac{E_{f}(\rho^{\otimes n})}{n}.$

(6). $E_{c}(\rho^{\otimes k}) = kE_{c}(\rho)$, where $k =1
,2,\cdots$.

Generally, by using a similar approach to proof of fact (6), we can
also have that

(7). $E_{d}(\rho^{\otimes k})=kE_{d}(\rho)$, $k = 1,2,\cdots$.

(8).  If $\rho$ is a product of two pure state
$\out{\psi}{\psi}\otimes\out{\phi}{\phi}$, then
\begin{eqnarray*} E_{c}(\out{\psi}{\psi}\otimes\out{\phi}{\phi}) = \
E_{d}(\out{\psi}{\psi}\otimes\out{\phi}{\phi}) = E(\ket{\psi})+E(\ket{\phi}).
\end{eqnarray*}

However, when $\rho$ and $\sigma$ are not the pure states, we do not know whether the additivity of $E_{c}$ and $E_{d}$ hold.

In \cite{Horodecki}, Professors P.~Horodecki, R.~Horodecki and
M.~Horodecki introduced the following important notion:

\begin{definition} Let $\cH=\bigotimes_{l=1}^{m}\cH_{l}$. We say that two pure states $\out{\psi}{\psi}$ and
$\out{\phi}{\phi}$ on $\cH$ are \emph{k-locally orthogonal}, if there exist some $k$
subsystems $\cH_{i_{1}},\ldots,\cH_{i_{k}}$ such that
$$
\Tr{(\out{\psi}{\psi})_l(\out{\phi}{\phi})_{l}}=0,\  l=i_{1},\ldots,i_{k},$$ where
$(\out{\psi}{\psi})_l=Tr_{\bigotimes_{j=1,j\neq l}^{m}\cH_{j}}\out{\psi}{\psi}$.

Moreover, the set of pure
states $\{\out{\psi_i}{\psi_i}\}_{i=1}^{K} $ is said to be
\emph{locally orthogonal} if $\{\out{\psi_i}{\psi_i}\}_{i=1}^{K} $ can be ordered in the
sequence
$\{\out{\psi_{i_{l}}}{\psi_{i_{l}}}\}_{i_{l}=i_1}^{i_{K}}$
such that for any $1\leq l \leq K$, the pure state $\out{\psi_{i_{l}}}{\psi_{i_{l}}}$ and the pure state $\out{\psi_{i_{n}}}{\psi_{i_{n}}}$ are
\emph{1-locally orthogonal} on the same subsystem whenever $n > l$.
\end{definition}

They proved also the following conclusion \cite{Horodecki}:

\begin{prop}\label{th:Horodecki} Let $\cH=\cH_{A}\bigotimes \cH_{B}$, $\rho$ be a quantum state on $\cH$.
If $\rho$ is composed of the locally orthogonal pure state ensemble $\{\out{\psi_{i}}{\psi_{i}}\}_{i=1}^K$ with probability distribution $p=(p_i)$ such that $\rho =
\sum_{i=1}^{K}p_{i}\out{\psi_{i}}{\psi_{i}}$, then
\begin{enumerate}[(i)]
\item\label{1} $E_{f}(\rho) = \sum_{i}p_{i}E_{f}(\out{\psi_{i}}{\psi_{i}})$,
\item\label{2} $E_{d}(\rho) = E_{f}(\rho) = E_{c}(\rho)$.
\end{enumerate}
\end{prop}

Proposition \ref{th:Horodecki} showed that the locally orthogonal pure states can be
distinguished by LOCC super-operators without destroying them.

\section{The Local orthogonality between mixed states}

In order to state our results, firstly, we need to extend the local orthogonality to the general quantum states.

\begin{definition}
Let $\cH=\bigotimes_{l=1}^{m}\cH_{l}$, We say that two quantum states $\rho$ and $\sigma$ on $\cH$ are $k$-locally orthogonal, if there exist some $k$
subsystems $\cH_{i_{1}},\ldots,\cH_{i_{k}}$, such that
$$
\Tr{(\rho)_l(\sigma)_{l}}=0,\  l=i_{1},\ldots,i_{k},$$
where
$(\rho)_l= Tr_{\bigotimes_{j=1,j\neq l}^{m}\cH_{j}}\rho$.

Moreover, the set of quantum states $\{\rho^{(a)}\}_{a\in\Sigma}$ on $\cH$ is said to be
\emph{locally orthogonal}, if $\{\rho^{(a)}\}_{a\in\Sigma}$ can be ordered in the
sequence $\{\rho^{(a_{1})},\rho^{(a_{2})},\ldots,\rho^{(a_{K})}\}$
such that for each $1\leq q \leq K$, the quantum state $\rho^{(a_{q})}$ and
the quantum state
$\rho^{(a_{n})}$ is 1-locally
orthogonal on the same subsystem whenever $n > q$, where $|\Sigma| = K$.
\end{definition}

In order to prove our main results, we need also the following
Proposition, which can be proved easily by the definition of local
orthogonality:

\begin{prop}\label{th:prop-1} Let $\cH=\cH_{A}\bigotimes \cH_{B}$, $\rho_1$ and $\rho_2$ are two locally orthogonal quantum states on $\cH$. Then $\langle Tr_{A}(\rho_{1}), Tr_{A}(\rho_{2})\rangle = 0$ or $
\langle Tr_{B}(\rho_{1}), Tr_{B}(\rho_{2})\rangle = 0$. Moreover, if
$\cH_{A}=\bigotimes_{i=1}^{m}\cH_{i}^{(A)}$,
$\cH_{B}=\bigotimes_{j=1}^{n}\cH_{j}^{(B)}$, $\langle
Tr_{A}(\rho_{1}), Tr_{A}(\rho_{2})\rangle = 0$, then
$Tr_{A}(\rho_{1})$ and $Tr_{A}(\rho_{2})$ are locally
orthogonal on $\cH_{B}$. If $ \langle Tr_{B}(\rho_{1}), Tr_{B}(\rho_{2})\rangle =
0$, then $Tr_{B}(\rho_{1})$ and $Tr_{B}(\rho_{2})$ are locally
orthogonal on $\cH_{A}$, too.
\end{prop}

\section{Main results}

In this section, we show that if the quantum state $\rho$ is composed of the locally orthogonal quantum state ensemble $\{\rho_{a}\}_{a\in\Sigma}$ with probability distribution $p=(p_a)$ such that $\rho = \sum_{a\in\Sigma}p_{a}\rho_{a}$, then the
entanglement of $\rho$ can be decomposed into the entanglement of
$\{\rho_{i}\}$ without losing them.

\begin{lem}\label{lem:lem-2}
Let $\cH = \cH_{A}\bigotimes \cH_{B}$, $\rho, \sigma$ be quantum
states on $\cH$. If the entanglement cost of quantum product state
$\rho\otimes\sigma$ is additive, that is $E_{c}(\rho\otimes\sigma) =
E_{c}(\rho) + E_{c}(\sigma),$ then
\begin{eqnarray*}
\lim_{n_1, n_2\rightarrow\infty}\frac{E_{f}(\rho^{\otimes n_1})+
E_{f}(\sigma^{\otimes n_2})}{n_1 + n_2} = \lim_{n_1,
n_2\rightarrow\infty}\frac{E_{f}(\rho^{\otimes
n_1}\otimes\sigma^{\otimes n_2})}{n_1 + n_2},
\end{eqnarray*}
where $n_1, n_2\in\mathbb{N}$, and there exists a positive number
$p>1$ such that $\lim_{n_1, n_2\rightarrow \infty}\frac{n_2}{n_1} =
p.$
\end{lem}

\begin{proof}
Consider the additive of entanglement cost, by the property (5) of
the entanglement, we have
\begin{eqnarray*}
\lim_{n\rightarrow \infty}\frac{E_{f}(\rho^{\otimes n})+
E_{f}(\sigma^{\otimes n})}{n} = \lim_{n\longrightarrow
\infty}\frac{E_{f}((\rho\otimes\sigma)^{\otimes n})}{n} <\infty.
\end{eqnarray*}
Therefore, for each $\varepsilon_1>0$, there exists a number
$N_1\in\mathbb{N}$, such that for any $n> N_1$, we have
\begin{eqnarray*}
0 < \frac{E_{f}(\rho^{\otimes n})+ E_{f}(\sigma^{\otimes n}) -
E_{f}((\rho\otimes\sigma)^{\otimes n})}{n} < \varepsilon_1,
\end{eqnarray*}
and for each $\varepsilon_2>0$, there exists a number
$N_2\in\mathbb{N}$, such that for any $n_{2} > n_{1} > N_1$, we have
\begin{eqnarray*}
|\frac{E_{f}(\rho^{\otimes n_1})}{n_1} - \frac{E_{f}(\rho^{\otimes
n_2})}{n_2}| < \varepsilon_2.
\end{eqnarray*}

Now, let us assume that there exists a positive number $a>0$ such
that
\begin{eqnarray*}
\lim_{n_1, n_2\rightarrow\infty}\frac{E_{f}(\rho^{\otimes n_1})+
E_{f}(\sigma^{\otimes n_2}) - E_{f}(\rho^{\otimes
n_1}\otimes\sigma^{\otimes n_2})}{n_1 + n_2} = a,
\end{eqnarray*}
where $n_1, n_2\in\mathbb{N}$, and there exists a positive number
$p>1$ such that $\lim_{n_1, n_2\rightarrow \infty}\frac{n_2}{n_1} =
p(n_1<n_2),$ then for each $\varepsilon_3>0$, there exists a number
$N_3\in\mathbb{N}$, such that for any $n_{2} > n_{1} > N_1$, we have
\begin{eqnarray*}
a - \varepsilon_3 < \frac{E_{f}(\rho^{\otimes n_1})+
E_{f}(\sigma^{\otimes n_2}) - E_{f}(\rho^{\otimes
n_1}\otimes\sigma^{\otimes n_2})}{n_1 + n_2} < a + \varepsilon_3,
\end{eqnarray*}
and for each $\varepsilon_4>0$, there exists a number
$N_4\in\mathbb{N}$, such that for any $n_{2} > n_{1} > N_4$, we have
\begin{eqnarray*}
p - \varepsilon_4 < \frac{n_2}{n_1} < p + \varepsilon_4.
\end{eqnarray*}
Therefore, when $N = \max\{N_1, N_2, N_3, N_4\}$, for each $n_2 >
\min\{n_1, n_2 - n_1\} > N,$ it follows that
\begin{eqnarray*}
-(a + \varepsilon_3) &<& \frac{E_{f}(\rho^{\otimes n_2})+
E_{f}(\sigma^{\otimes n_2}) - E_{f}((\rho\otimes\sigma)^{\otimes
n_2})}{n_1 + n_2} - \frac{E_{f}(\rho^{\otimes n_1})+
E_{f}(\sigma^{\otimes n_2}) - E_{f}(\rho^{\otimes
n_1}\otimes\sigma^{\otimes n_2})}{n_1 + n_2}\\
&=& \frac{E_{f}(\rho^{\otimes n_2}) - E_{f}(\rho^{\otimes n_1}) -
E_{f}((\rho\otimes\sigma)^{\otimes n_2}) + E_{f}(\rho^{\otimes
n_1}\otimes\sigma^{\otimes n_2})}{n_1 + n_2} \\
&<& -(a - \frac{(p +
\varepsilon_4)\varepsilon_1}{1 + p - \varepsilon_4} -
\varepsilon_3),
\end{eqnarray*}
note that
\begin{eqnarray*}
\frac{E_{f}(\rho^{\otimes (n_2-n_1)})}{n_1 + n_2} - \varepsilon_2 <
\frac{E_{f}(\rho^{\otimes n_2}) - E_{f}(\rho^{\otimes n_1})}{n_1 +
n_2} < \frac{E_{f}(\rho^{\otimes (n_2-n_1)})}{n_1 + n_2} +
\varepsilon_2,
\end{eqnarray*}
then we obtain that
\begin{eqnarray*}
-(a + \varepsilon_2 + \varepsilon_3) &<& \frac{E_{f}(\rho^{\otimes
(n_2-n_1)}) - E_{f}((\rho\otimes\sigma)^{\otimes n_2}) +
E_{f}(\rho^{\otimes
n_1}\otimes\sigma^{\otimes n_2})}{n_1 + n_2} \\
&<& -(a - \frac{(p + \varepsilon_4)\varepsilon_1}{1 + p -
\varepsilon_4} - \varepsilon_2 - \varepsilon_3).
\end{eqnarray*}
But we choose $\varepsilon_1, \varepsilon_2, \varepsilon_3,
\varepsilon_4 > 0$ small enough so that $\varepsilon_1 =
\varepsilon_2 = \varepsilon_3 < a/3$ and $\varepsilon_4 < 1/2$, then
\begin{eqnarray*}
E_{f}(\rho^{\otimes (n_2-n_1)}) + E_{f}(\rho^{\otimes
n_1}\otimes\sigma^{\otimes n_2}) < E_{f}(\rho^{\otimes (n_2-n_1)}
\otimes \rho^{\otimes n_1}\otimes\sigma^{\otimes n_2}) =
E_{f}((\rho\otimes\sigma)^{\otimes n_2}),
\end{eqnarray*}
this contradicts the property (1) of entanglement. Hence, we have
\begin{eqnarray*}
\lim_{n\rightarrow\infty}\frac{E_{f}(\rho^{\otimes n_1})+
E_{f}(\sigma^{\otimes n_2}) - E_{f}(\rho^{\otimes
n_1}\otimes\sigma^{\otimes n_2})}{n_1 + n_2} = 0,
\end{eqnarray*}
which completes the proof.
\end{proof}

\begin{thrm}\label{th:mainresult-1} Let $\cH = \cH_{A}\bigotimes \cH_{B}$, $\rho$ be a quantum state on
 $\cH$. If $\rho$ is composed of the locally orthogonal quantum ensemble $\{\rho_{a}\}_{a\in\Sigma}$ with probability distribution $p=(p_a)$ such that $\rho = \sum_{a\in\Sigma}p_{a}\rho_{a}$, then
\begin{eqnarray*}
E_{f}(\rho) = \sum_{a\in\Sigma}p_{a}E_{f}(\rho_{a}).
\end{eqnarray*}

Moreover, if the entanglement cost $E_{c}$ is additive for the
quantum product state $\bigotimes_{a\in\Sigma}\rho_{a}$, that is,
$E_{c}(\bigotimes_{a\in\Sigma}\rho_{a}) =\sum_{a\in\Sigma}
E_{c}(\rho_{a})$, then
\begin{eqnarray*}
E_{c}(\rho) = \sum_{a\in\Sigma}p_{a}E_{c}(\rho_{a}).
\end{eqnarray*}
\end{thrm}

\begin{proof} If $\rho=p_{1}\rho_{1}+p_{2}\rho_{2}$, where $p = (p_1, p_2)$ is a probability distribution, the quantum states $\rho_{1}, \rho_{2}$ are locally orthogonal. By the definition of locally
orthogonal, we know that
\begin{eqnarray*}
\langle Tr_{A}(\rho_{1}), Tr_{A}(\rho_{2})\rangle = 0,\ \ or \ \
\langle Tr_{B}(\rho_{1}), Tr_{B}(\rho_{2})\rangle = 0.
\end{eqnarray*}

Without lost generality, we assume that $\langle Tr_{A}(\rho_{1}),
Tr_{A}(\rho_{2})\rangle = 0,$ then by Proposition \ref{th:prop-1},
$Tr_{A}(\rho_{1})$ and $Tr_{A}(\rho_{2})$ are orthogonal, it implies
that there are subspaces $\cV_{i}^{B} \subseteq \cH_{B}$ such that
$\rho_{i} \in \D(\cH_{A}\bigotimes \cV_{i}^{B})$ and
$\cV_{2}^{B}\subseteq (\cV_{1}^{B})^{\perp}$. The inequality
$E_{f}(\rho) \leq p_{1}E_{f}(\rho_{1}) + p_{2}E_{f}(\rho_{2})$
follows from convexity of $E_{f}$. The reverse inequality is a
consequence of the monotonicity property (2) of the entanglement
applied to the maps
\begin{eqnarray*}
\cM^{(i)}_{loc}(\rho) = 1^{A}\otimes\pi^{B}_{i} \ \rho \ 1^{A}\otimes\pi^{B}_{i}, \ \ i = 1,2,3,
\end{eqnarray*}
where $\cV_{3}^{B} = 1_{B} - (\cV_{1}^{B} + \cV_{2}^{B})$, and $\pi^{B}_{i}$ are the projectors onto $\cV_{i}^{B}$, respectively.
It follows that
\begin{eqnarray*}
\sum_{a=1,2}p_{a} E_{f}(\rho_{a}) = \sum^{3}_{i=1}q_{i}E_{f}(q_{i}^{-1}\cM^{(i)}_{loc}(\rho))
\leq E_{f}(\rho),
\end{eqnarray*}
with $q_{i} = Tr[\cM^{(i)}_{loc}(\rho)].$

Repeatedly, when the set of quantum states $\{\rho_{a}\}$ is
locally orthogonal, then
\begin{eqnarray*}
E_{f}(\rho) = \sum_{a}p_{a}E_{f}(\rho_{a}).
\end{eqnarray*}

Next, regarding the entanglement cost $E_{c}(\rho)$, let the quantum
state $\rho^{\otimes n} = \sum_{t\in\Sigma^{n}}p_{t}\rho_{t}$ such
that all quantum states $\rho_{t} = \rho_{a_1} \otimes \rho_{a_2}
\otimes \cdots \otimes \rho_{a_n}$, where $t = a_1 \cdots a_n$.
Then, by property of locally orthogonal quantum states
$\{\rho_{t}\}$, we have
\begin{eqnarray*}
E_{f}(\rho^{\otimes n}) = \sum_{t\in\Sigma^{n}}p_{t}E_{f}(\rho_{t}).
\end{eqnarray*}
And, for any positive real number $\varepsilon$, by the property (5)
of the entanglement and Lemma~\ref{lem:lem-1}, we have
\begin{eqnarray*}
E_{c}(\rho) &=& \lim_{n\rightarrow\infty} \frac{E_{f}(\rho^{\ot n})}{n} =
\lim_{n\rightarrow\infty} \frac{\sum_{t\in\Sigma^{n}}p_{t}E_{f}(\rho_{t})}{n}\\
&=& \lim_{n\rightarrow\infty} [\sum_{s\in
T_{\varepsilon}^{n}}\{(\prod_{a\in \Sigma}p_{a}^{n_{a,s}})
\frac{E_{f}(\rho_{s})}{n}\}],
\end{eqnarray*}
where $\sum_{a\in \Sigma} n_{a,s} = n$ for $s\in
T_{\varepsilon}^{n}.$  Also, by Lemma \ref{lem:lem-2}, if we have
$E_{c}(\bigotimes_{a\in\Sigma}\rho_{a}) =\sum_{a\in\Sigma}
E_{c}(\rho_{a}),$ then
\begin{eqnarray*}
\lim_{n\rightarrow\infty}\frac{E_{f}(\rho_{s})}{n} =
\lim_{n\rightarrow\infty}\frac{\sum_{a\in\Sigma}E_{f}(\rho_{a}^{\bigotimes
n_{a,s}})}{n}.
\end{eqnarray*}

Therefore
\begin{eqnarray*}
E_{c}(\rho) &=& \lim_{n\rightarrow\infty} [\sum_{s\in
T_{\varepsilon}^{n}}\{(\prod_{a\in\Sigma}p_{a}^{n_{a,s}})
\sum_{a\in\Sigma}\frac{(n_{a,s})E_{f}(\rho_{a}^{\bigotimes n_{a,s}})}{n(n_{a,s})}\}] \\
&=& \lim_{n\rightarrow\infty} [\sum_{s\in
T_{\varepsilon}^{n}}\{(\prod_{a\in\Sigma}p_{a}^{n_{a,s}})
\sum_{a\in\Sigma}\frac{(p_{a})E_{f}(\rho_{a}^{\bigotimes n_{a,s}})}{(n_{a,s})}\}]\\
& & + \lim_{n\rightarrow\infty} [\sum_{s\in
T_{\varepsilon}^{n}}\{(\prod_{a\in\Sigma}p_{a}^{n_{a,s}})
\sum_{a\in\Sigma}\frac{\{(n_{a,s}/n) -
p_{a}\}E_{f}(\rho_{a}^{\bigotimes n_{a,s}})}{(n_{a,s})}\}].
\end{eqnarray*}
It follows from $|\frac{n_{a,s}}{n} - p_{a}|\leq \frac{\varepsilon
\log_{p_{a}}2}{|\Sigma|} \rightarrow 0$\ as \ $
\varepsilon\rightarrow 0$ and $n\rightarrow\infty$ that
\begin{eqnarray*}
\lim_{n\rightarrow\infty} [\sum_{s\in
T_{\varepsilon}^{n}}\{(\prod_{a\in\Sigma}p_{a}^{n_{a,s}})
\sum_{a\in\Sigma}\frac{\{(n_{a,s}/n) -
p_{a}\}E_{f}(\rho_{a}^{\bigotimes n_{a,s}})}{(n_{a,s})}\}] = 0,
\end{eqnarray*}
and, by \ $\sum_{s\in
T_{\varepsilon}^{n}}(\prod_{a\in\Sigma}p_{a}^{n_{a,s}})\rightarrow
1$ as $\ n\rightarrow\infty$, we have
\begin{eqnarray*}
\lim_{n\rightarrow\infty} [\sum_{s\in
T_{\varepsilon}^{n}}\{(\prod_{a\in\Sigma}p_{a}^{n_{a,s}})
\sum_{a\in\Sigma}\frac{(p_{a})E_{f}(\rho_{a}^{\bigotimes
n_{a,s}})}{(n_{a,s})}\}] = \sum_{a\in\Sigma}p_{a} E_{c}(\rho_{a}).
\end{eqnarray*}
\end{proof}

For the distillable entanglement, we also hope to have the same decomposition under the condition of locally orthogonal. But it is hard to find certain conditions for holding the same result as equality, we only have a weak result as inequality.

Firstly, we need the following lemma for our result.

\begin{lem}\label{lem:lem-3}
Consider any tensor product quantum states $\rho\otimes\varrho$ of
the quantum system composed from two subsystems, then
\begin{eqnarray*}
E_{d}(\rho\otimes\varrho)\geq E_{d}(\rho) + E_{d}(\varrho).
\end{eqnarray*}
\end{lem}

\begin{proof}
Assume that $E_{d}(\rho) = \alpha$, $E_{d}(\sigma) = \beta$, then
there exists sequences $\{\Phi_{n}, \Psi_{n} : n \in \mathbb{N}\}$
of LOCC super-operators such that
\begin{eqnarray*}
\lim_{n\rightarrow\infty}F(\Phi_{n}(\rho^{\otimes n}), \tau^{\ot\lfloor \alpha n\rfloor}) = 1, \ \ \ \  \lim_{n\rightarrow\infty}F(\Psi_{n}(\sigma^{\otimes n}), \tau^{\ot\lfloor \beta n\rfloor}) = 1,
\end{eqnarray*}
where $\tau = \frac{1}{2}(\ket{00}+\ket{11})(\bra{00}+\bra{11}).$
Therefore, we can construct a sequence $\{\Xi_{n} : n \in \mathbb{N}\}$ of LOCC super-operators such that
\begin{eqnarray*}
\lim_{n\rightarrow\infty}F(\Xi_{n}((\rho\otimes \sigma)^{\otimes n}), \tau^{\ot(\lfloor \alpha n\rfloor+\lfloor \beta n\rfloor)}) = 1.
\end{eqnarray*}

Firstly, if at least one of the two numbers $\alpha, \beta$ are
integers, note that
\begin{eqnarray*}
\lfloor \alpha n\rfloor+\lfloor \beta n\rfloor = \lfloor (\alpha+\beta) n\rfloor,
\end{eqnarray*}
and $E_{d}(\rho\otimes \sigma)$ is the supremum, we have
$E_{d}(\rho\otimes \sigma) \geq E_{d}(\rho) + E_{d}(\sigma)$ as
required.

Next, if all numbers $\alpha, \beta$ are not integers, then for
$\varepsilon>0$ small enough so that $0<\alpha+\beta-\varepsilon$,
there exists an integer $N$ such that for all $n\geq N$,
\begin{eqnarray*}
\lfloor (\alpha+\beta-\varepsilon) n\rfloor \leq \lfloor \alpha n\rfloor+\lfloor \beta n\rfloor \leq \lfloor (\alpha+\beta) n\rfloor.
\end{eqnarray*}
Also, for all $n$, we know
$Tr_{(\cC^{\{0,1\}}\ot\cC^{\{0,1\}})^{\otimes\gamma_{n}}}$ are LOCC
super-operators, where $\gamma_{n} = \lfloor \alpha n\rfloor+\lfloor
\beta n\rfloor - \lfloor (\alpha+\beta-\varepsilon) n\rfloor,$ thus,
\begin{eqnarray*}
\lim_{n\rightarrow\infty}F(Tr_{(\cC^{\{0,1\}}\ot\cC^{\{0,1\}})^{\otimes\gamma_{n}}}\Xi_{n}((\rho\otimes
\sigma)^{\otimes n}), \tau^{\ot(\lfloor (\alpha+\beta-\varepsilon)
n\rfloor)}) = 1.
\end{eqnarray*}
Therefore, by the arbitrariness of $\varepsilon$, it follows that $E_{d}(\rho\otimes \sigma) \geq E_{d}(\rho) + E_{d}(\sigma)$.
\end{proof}

\begin{thrm}\label{th:mainresult-2}
Let $\cH = \cH_{A}\bigotimes \cH_{B}$, $\rho$ be a quantum state on
 $\cH$. If $\rho$ is composed of the locally orthogonal quantum ensemble $\{\rho_{a}\}_{a\in\Sigma}$ with probability distribution $p=(p_a)$ such that $\rho = \sum_{a\in\Sigma}p_{a}\rho_{a}$, then
\begin{eqnarray*}
E_{d}(\rho) \geq \sum_{a\in\Sigma}p_{a}E_{d}(\rho_{a}).
\end{eqnarray*}
\end{thrm}

\begin{proof}
Consider the distillable entanglement of quantum state $\rho =
\sum_{a\in\Sigma}p_{a}\rho_{a}$. By definition of the distillable
entanglement and Lemma~\ref{lem:lem-1}, $E_{d}(\rho)$ is the
supremum over all real numbers $\alpha\geq 0$ for which there exists
a sequence of LOCC super-operators $\{\Phi_{n} : n \in
\mathbb{N}\}$, such that
\begin{eqnarray*}\label{eq:anotherversion-3}
\lim_{n\rightarrow\infty}F(\Phi_{n}(\rho^{\otimes n}),
\tau^{\ot\lfloor \alpha n\rfloor})  =
\lim_{n\rightarrow\infty}F(\Phi_{n}(\rho_{T_{\varepsilon}^{n}}),
\tau^{\ot\lfloor \alpha n\rfloor})= 1,
\end{eqnarray*}
where $\tau = \frac{1}{2}(\ket{00}+\ket{11})(\bra{00}+\bra{11}),$
and the set $T_{\varepsilon}^{n}$ is the typical set for the
probability distribution $p$. It follows from the observation that
the state $\tau^{\ot n}$ is a pure state for all positive numbers
$n\geq1$ that
\begin{eqnarray*}\label{eq:anotherversion-3}
\lim_{n\rightarrow\infty}(F(\Phi_{n}(\rho^{\otimes n}), \tau^{\ot\lfloor \alpha n\rfloor}))^{2} &=& \lim_{n\rightarrow\infty}(F(\Phi_{n}(\rho_{T_{\varepsilon}^{n}}), \tau^{\ot\lfloor \alpha n\rfloor}))^{2}\\
&=& \lim_{n\rightarrow\infty}[\sum_{s\in
T_{\varepsilon}^{n}}\{(\prod_{a\in\Sigma}p_{a}^{n_{a,s}})(F(\Phi_{n}(\rho_{s}), \tau^{\ot\lfloor \alpha n\rfloor}))^{2}\}].
\end{eqnarray*}
Also, because the locally orthogonal set can be distinguished by using LOCC super-operators
without destroying them \cite{Horodecki}, there exists a sequence of
LOCC super-operators $\{\Phi_{n} : n \in \mathbb{N}\}$ for all
quantum states
$\rho_{s}$, such that
\begin{eqnarray*}
\lim_{n\rightarrow\infty}F(\Phi_{n}(\rho_{s}), \tau^{\ot\lfloor \beta n\rfloor}) = 1,
\end{eqnarray*}
where $\beta =
\lim_{n\rightarrow\infty}\frac{E_{d}(\bigotimes_{a\in\Sigma}\rho_{a}^{\otimes
n_{a,s}})}{n}$, thus
\begin{eqnarray*}
\lim_{n\rightarrow\infty}F(\Phi_{n}(\rho^{\otimes n}),
\tau^{\ot\lfloor \beta n\rfloor}) =
\lim_{n\rightarrow\infty}\sqrt{\sum_{s\in
T_{\varepsilon}^{n}}\{(\prod_{a\in\Sigma}p_{a}^{n_{a,s}})(F(\Phi_{n}(\rho_{s}),
\tau^{\ot\lfloor \beta n\rfloor}))^{2}\}} = 1.
\end{eqnarray*}

Therefore, by Lemma~\ref{lem:lem-3}, we have
\begin{eqnarray*}
E_{d}(\rho) \geq
\lim_{n\rightarrow\infty}\frac{E_{d}(\bigotimes_{a\in\Sigma}\rho_{a}^{\otimes
n_{a,s}})}{n} \geq \sum_{a}p_{a}E_{d}(\rho_{a})
\end{eqnarray*}
as required.
\end{proof}

\section{Example of main result}

Finally, we present an interesting example to show that the conditions of Theorem~\ref{th:mainresult-1} and Theorem~\ref{th:mainresult-2} are existence.

\begin{exam}
Consider $\cH_{A} = \cC^{3}, \cH_{B} = \cC^{6}$, and the subspace
$\cH_{V}\subseteq \cH_{A}\otimes \cH_{B}$ spanned by
\begin{eqnarray*}
\ket0_{V} \equiv \frac{1}{2}(\ket1_{A}\ket2_{B} + \ket2_{A}\ket1_{B} + \sqrt{2}\ket0_{A}\ket3_{B}),\\
\ket1_{V} \equiv \frac{1}{2}(\ket2_{A}\ket0_{B} + \ket0_{A}\ket2_{B} + \sqrt{2}\ket1_{A}\ket4_{B}),\\
\ket2_{V} \equiv \frac{1}{2}(\ket0_{A}\ket1_{B} + \ket1_{A}\ket0_{B} + \sqrt{2}\ket0_{A}\ket5_{B}).
\end{eqnarray*}
Then for each $\rho_{V}\in D(\cH_{V})$ and $\sigma\in D(\cH_{A}\otimes
\cH_{B})$, it follows from \cite{Vidal} that $E_{f}(\rho_{V}\otimes\sigma) =
E_{f}(\rho_{V}) + E_{f}(\sigma)$. This implies that
\begin{eqnarray}\label{eq:anotherversion-1}
E_{c}(\rho_{V}\otimes\sigma) = E_{c}(\rho_{V}) + E_{c}(\sigma).
\end{eqnarray}

In this condition, let us consider the quantum states
$\rho_{V}=\sum_{a\in\Sigma_{V}}p_{a}\ket{\phi_{a}}_{V}\bra{\phi_{a}},\
\sigma=\sum_{b\in\Sigma_{AB}}q_{b}\ket{\psi_{b}}_{AB}\bra{\psi_{b}}$
for any probability distributions $p = (p_{a})$ and $q = (q_{b})$,
where
\begin{eqnarray*}
\ket{\phi_{a}}_{V} &=& r_{a}\ket1_{V} + \sqrt{1-r_{a}^{2}}\ket2_{V} \ \ (r_{a} \in [-1,1]), \\
\ket{\psi_{b}}_{AB} &=& (s_{0,b}\ket0_{A} + s_{1,b}\ket1_{A} +
s_{2,b}\ket2_{A})\ket3_{B},\ \ and\ \ s_{0,b}^{2} + s_{1,b}^{2} +
s_{2,b}^{2} = 1.
\end{eqnarray*}
Then we know the set of quantum states $\{\rho_{V}, \sigma\}$ is
locally orthogonal. Consequently, the conditions of
Theorem~\ref{th:mainresult-1} and Theorem~\ref{th:mainresult-2} can be satisfied.
\end{exam}


\subsection*{Acknowledgements} This  project is supported by Research Fund, Kumoh National Institute of Technology.




\end{document}